\documentstyle[times,pramana,epsf,epsfig,psfig,subeqn,floats]{ias}
\def\Journal#1#2#3#4{{#1} {\bf #2}, #3 (#4)}

\def\NPB{{\em Nucl. Phys.} B}
\def\PLB{{\em Phys. Lett.}  B}
\def\PR{\em Phys. Reports}
\def\EUJ{{\em Eur. Phys. J.} C}
\def\PRL{\em Phys. Rev. Lett.}
\def\PRD{{\em Phys. Rev.} D}
\def\ZPC{{\em Z. Phys.} C}
\def\PR{{\em Phys. Reports}}
\def\RMP{\em Rev. of Modern Physics}
\def\JHEP{\em Journal of High Energy Physics}
\def\pramana{\em Pramana}
\def\be{\begin{equation}}
\def\ee{\end{equation}}
\def\bea{\begin{eqnarray}}
\def\eea{\end{eqnarray}}
\def\to{\rightarrow}

\def\eplem{\mbox{$e^+e^-$}}
\def\gamgam{\mbox{$\gamma \gamma$}}
\def\mh{\mbox{$m_h$}}
\def\mhzer{\mbox{$m_{h_0}$}}
\def\hzer{\mbox{${h_0}$}}
\def\Hzer{\mbox{${H_0}$}}
\def\mHzer{\mbox{$m_{H_0}$}}
\def\mt{\mbox{$m_t$}}
\def\mz{\mbox{$m_Z$}}
\def\ma{\mbox{$m_A$}}
\def\mhc{\mbox{$m_{H^\pm}$}}

\def\rpv{\mbox{$R_p \hspace{-1em}/\;\:$}}
\def\rts{\mbox{$\sqrt{s}$}}
\def\lsim{\:\raisebox{-0.5ex}{$\stackrel{\textstyle<}{\sim}$}\:}
\def\gsim{\:\raisebox{-0.5ex}{$\stackrel{\textstyle>}{\sim}$}\:}

\begin{document}

\mark{{Higgs and SUSY searches...}{R M Godbole}}
\title{Higgs and SUSY searches at future Colliders}

\author{R M GODBOLE}
\address{Centre for Theoretical Studies, Indian Institute of Science, 
Bangalore 560012}
\keywords{Higgs;Supersymmetry;Colliders.}
\pacs{12.15.-y;14.80.-j;14.80.Cp;14.80.Bn;14.80.Ly}
\abstract
{In this talk, I discuss some aspects of Higgs searches at future colliders, 
particularly comparing and contrasting the capabilities of  LHC and Next 
Linear Collider (NLC), 
including the aspects of Higgs searches in supersymmetric theories. I will 
also discuss how the search and study of sparticles other than the Higgs can 
be ysed to give information about the parameters of the Minimal 
Supersymmetric Standard Model (MSSM).}
\maketitle
\section{Introduction}

	The SM has been tested to an unprecedented accuracy over the past few 
decades culminating in the precision measurement at LEP as well as the 
observation of WW production at LEP-II~\cite{1}. The agreement   
with the SM predictions of the precision measurements at LEP  as well as that 
of the WW cross-section at LEP-II proves that the SM is 
described by a renormalizable, $SU(2) \times U(1)$ gauge field theory. 
The renormalizability of the theory requires 
the Spontaneous Symmetry Breaking (SSB) and , in the currently accepted 
theoretical dogma, Higgs mechanism. However, the direct search for the 
elusive Higgs has only resulted in lower limits which give 
$m_{h} > 89.7$ GeV~\cite{LEPHWG}\footnote {Indications from the latest 
results from the LEP collaborations~\cite{Moriond} is that this limit 
will creep upto $108$ GeV.}.

Thus at present we have no `direct' proof that Higgs mechanism is `the' 
mechanism for the SSB. Further, quantum field theories with fundamental 
scalars require some mechanism to stabilize  the mass of the Higgs
\mh\  around the scale 
of the Electroweak (EW) symmetry breaking. Unless the TeV scale gravity 
obviates the problem itself ~\cite{2}, Supersymmetry is the best available 
option for the purpose~\cite{3}. Since the `Raison d'etre' for future 
colliders is to establish the mechanism of the SSB, it is clear that 
`Higgs and Supersymmetry search' is the most important aspect of physics 
at the next generation colliders.

	In view of the importance of these two searches there exist a 
large number of discussions of the phenomenological and experimental 
possibilities in the context of future colliders
in literature~\cite{cms1,21,6,5,20,9P,4,7P,6P,8P}. In this talk,
I will focus on some of the recent developments and questions in the 
Higgs search, {\it viz.}

\begin{enumerate} 
\item In view of the LEP-II results what can Tevatron (and of course LHC) 
do for Higgs search?
\item If `a' scalar is found at LHC how well can one decide that this 
scalar is `the' SM Higgs $h$?
\item What is the LHC reach for MSSM SUSY Higgs search?
\item If and how does NLC improve the situation?
\item What can $\gamma \gamma$ (and further $\mu^{+} \mu^{-}$) colliders 
do?
\end{enumerate}
As far as supersymmetry (SUSY) is concerned, it is clear that it is broken, 
albeit the breaking should be at the TeV scale if SUSY has anything to do with 
particle physics.        
However, there is no `real' understanding how SUSY is broken. So, what we 
really need to do with the SUSY particles (after finding them) is to measure  
their properties accurately and use them to study how SUSY is broken and learn 
something about the high scale physics from the way it is broken. It is by 
now clear that both the LHC and NLC can find TeV scale SUSY, if it exists. 
The emphasis of all the recent studies has been on how to test different 
models for high scale physics once SUSY is found.

	The future colliders that I would discuss would be mainly Tevatron 
(Run II,TEV33), LHC and the NLC. The specifications of a future linear collider 
are not yet completely finalized. The normally considered energies are
$\sqrt{s}_{\eplem} \lsim 500$ GeV, with luminosities  $ \sim~20 - 50\; 
{\rm fb}^{-1}$.  However, linear colliders with energies 
extended upto 2 TeV are considered and normally used integrated luminosities
for them are usually  scaled up to compensate for the $1/s$ factor in the 
cross-section. The technical feasibility of a 500 GeV linear collider is now 
established ~\cite {6}.  There are also discussions about constructing a 
$\gamma \gamma$ collider using backscattered laser photons from an \eplem\ 
machine. Particle production in real scattering of `real' high energy photons 
using backscattered laser photons has been observed for the first 
time~\cite {7}. In principle, in the $e^+e^-$ option of the 
Linear Colliders (LC's) one has to worry about new phenomena such 
as Beamstrahlung as well as the backgrounds caused by high energy 
$\gamma \gamma$ interactions~\cite{8}. 
But they are under control for \eplem\ colliders upto $\sqrt{s} =  1$ 
TeV~\cite{9}. The high degree of polarization possible at the LC's is 
particularly useful in precision studies of SUSY.
 
\section{Higgs  Search}
\subsection{Theoretical Mass Limits}
\subsubsection{SM Higgs}

In the SM only the couplings of Higgs with matter and gauge particles 
are predicted, but nothing much is known theoretically about its mass, 
apart from the limits. The  upper limit comes from triviality 
considerations~\cite {10,11} i.e.  by demanding that the Landau pole in 
Higgs self coupling $\lambda$ should not occur upto an energy scale $\Lambda$. 
The lower bound~\cite{12} 
comes from instability of vacuum under fluctuations. The latter is valid only 
if the Higgs content is minimal viz. a single doublet. Both the bounds 
depend on $m_t$. Given the fairly accurate knowledge on the mass of the top 
quark, this then gives predictions for both the bounds. The upper bound, 
in addition, depends on the uncertainties in nonperturbative dynamics which 
needs to be employed while analysing the large $\lambda$ region. 
\begin{figure}[htb]
\begin{center}
\vspace{-0.3cm}
%\mbox{\epsfig{file=../hadr13/limits.eps,height=60mm}}
\mbox{\epsfig{file=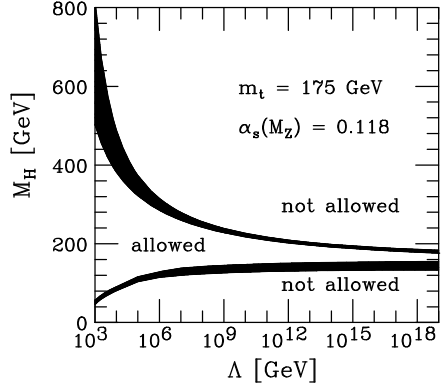,height=60mm}}
\caption{Theoretical bounds on \mh\ in the SM~\protect\cite{13}.
\label{fig1}}
\end{center}
\end{figure}
Fig.~\ref{fig1} shows the bounds obtained in a recent analysis~\cite{13}. This 
tells us that should we discover at Tevatron - run II some direct evidence 
of a Higgs with   mass $\sim  115$-120 GeV, we can take that as an indication 
that the desert between the Weak scale and the Planck scale is sure to be 
populated and  hence also that of possible new physics within the reach of LHC 
experiments.

\subsubsection{SUSY Higgs}

Before embarking upon a discussion of the search prospects of the 
SM Higgs at future colliders, let us also summarise  a few things 
about the predictions of the properties of the various scalars that exist
in Supersymmetric theories. The scalar sector is much richer in these
theories and there  are five scalars: three neutrals out of which two are 
CP even states: the lighter(heavier) one being denoted by  \hzer (\Hzer)
and  one CP odd state denoted by $A$ and a pair of charged Higgs bosons
$H^\pm$. The masses of all the scalars are not independent. They are
given in terms of two parameters, which can be chosen either to be
$\ma ,\tan \beta$ or $\mhc , \tan \beta$. Here $\tan \beta$ is the
ratio of the vacuum expectation values of the neutral members 
of the two Higgs doublets that exist in the MSSM.  As a result of the 
supersymmetry
these masses satisfy certain sum rules and hence inequalities at tree level:
\be
\mhzer \leq m_Z, \mHzer > m_Z, m_{H^\pm} > m_W, \mhzer < \mHzer, m_{H^\pm} .
\label{e1}
\ee
In the decoupling limit~\cite{14} ($m_A \to \infty $) one finds that, 
independent of $\tan \beta$,  all the four heavy scalars become degenerate and
infinitely heavy and the mass of the lightest scalar approaches the upper bound.
In this limit the couplings of the $h_0$ to matter fermions and the gauge 
bosons approach those of the SM Higgs $h$. 

%The interesting thing to note here
%is that all the masses $m_{H^\pm}, m_A, \mhzer$ can become large without some 
%self coupling becoming strong, unlike the case of the SM. 

The most interesting, of course, is the upper limit on the mass of the 
lightest neutral Higgs \hzer\  {\it viz.}  \mhzer.  These mass relations
receive large radiative corrections, due to the large mass of the top quark 
\mt . However, \mhzer\ is still bounded. Also note that the corrections 
will vanish in the limit of exact supersymmetry. The limits on the 
radiatively corrected scalar masses for the case of maximal mixing in 
\begin{figure}[htb]
\vspace{0.5cm}
\begin{center}
%\mbox{\epsfig{file=../hadr13/higloop.eps,height=45mm}}
\mbox{\epsfig{file=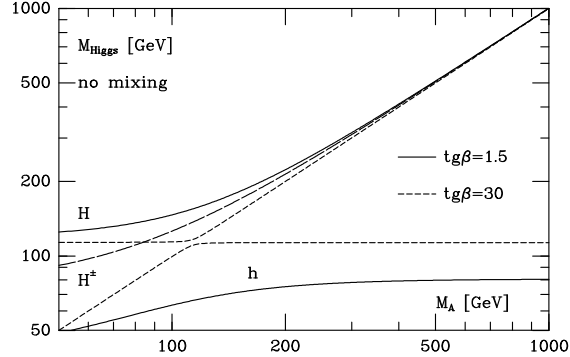,height=45mm}}
\vspace{-0.5cm}
\end{center}
\caption{Bounds on the masses of the scalars in the 
MSSM~\protect\cite{15}\label{fig2}}
\end{figure}
the stop sector are shown in Fig.~\ref{fig2}~\cite{15}.
It shows that the mass of the lightest scalar in MSSM  is bounded by 
$\sim 130 $ GeV even after it is radiatively corrected. This bound does 
get modified in the NMSSM~\cite{16,17,18}.  New results in this context are
the two loop calculations~\cite{heinemeyer} of the threshold corrections
to the effective quartic couplings of the Higgs potential.  These results show
that for all reasonable values of the model parameters, \mhzer\ is bounded by 
$\sim 150$ GeV. The couplings of \hzer\  do get modified to some extent by the 
radiative corrections, but the general features remain the same 
as the tree level results.

Not only are the scalars much more numerous in the MSSM, 
their decay patterns are  much more involved and depend on the
parameters of SUSY model, as both the exact masses and the couplings 
of the Higgses are dependent on these. Hence the phenomenology of the 
MSSM scalars is much richer and more complicated than the SM case. 
Again, calculation of various decay widths including the higher order 
corrections involving loops of sparticles has been 
done~\cite{abdel1,abdel2,spira}.

\subsection{Experimental limits}
The best direct limit on the mass of the Higgs  comes from the study 
$\eplem \to  Z^* h$ at LEP-II and  will soon reach 108 
GeV~\cite{Moriond}~\footnote{I have used here and later the updated numbers 
since the talk was presented.}.  This is already at the limit of the reach of 
LEP-II. It has been made possible due to the use of $ \eplem \to  
Z^* h \to \nu \bar \nu h$ and the excellent b-tagging achieved in the 
detectors at LEP-II. There exist also, the `indirect' bounds on \mh\
arrived at from the analysis of precision measurements from 
LEP ~\cite{19}. This gives a lower limit on \mh\ of 77 GeV with a $95 \%$
confidence level upper limit of 215 GeV. This then seems to be tantalizingly 
consistent with the predictions of the bound of $ 160 \pm  30$  GeV from the 
consistency of the SM. However, it should not be forgotten that new physics 
within 
the reach of LHC might change some of the theoretical predictions for the 
variables used in the precision data.  It should be thus borne in mind that
with the small expected improvements in the precision, EW data might be 
remain consistent with a somewhat higher upper limit on \mh. This is to 
say that it may be possible to relax somewhat the upper bound on \mh\ 
implied by the analysis~\cite{19}. The indirect limits on the  mass
of the lightest CP even neutral scalar in the MSSM  are very similar
to that on the SM Higgs, due to decoupling nature of SUSY. 
The latest (priliminary results on) mass limits for the SUSY Higgses
are : $ \mhzer > 88.3 {\rm GeV}; \ma > 88.4 {\rm GeV} $
and an absolute limit on $\tan \beta$ : $0.4 < \tan \beta < 4.1\;\; 
( 0.7 < \tan \beta < 1.8 )$ for no (large) mixing in 
the stop sector~\cite{Moriond}.

\subsection{Higgs search at the Hadronic colliders :}
The mass range for the search of the Higgs divides itself 
into two regions: i) $102 \leq \mh \lsim 150$ GeV, 
~ii) $\mh \gsim 150$ GeV. The lower limit in (i) is simply a reflection
of the current lower limit from `direct' searches for the Higgs. The upper 
limit of region (i) is decided by dominant decay modes of the
Higgs.  A large number of recent discussions~\cite{cms1,21,20,spira}, 
both theoretical 
and experimental, have concentrated on  Higgs search strategies in this 
mass range, for obvious reasons. This is the mass range preferred by
the `indirect' limits; this is also the mass range expected for
the lightest supersymmetric Higgs \hzer . It also happens to be the
mass range
that would be accessible at the Tevatron Run-II/TEV33 as far as the production
cross-sections are considered. From the point of view of a clean signal, this
happens to be the  most challenging range of the Higgs mass as the dominant 
branching ratio in this case is into the $b \bar b$ channel, where the QCD 
background is about three orders of magnitude higher than the signal. 

\subsubsection{SM Higgs}
Let us first discuss the  case of the search for the SM Higgs $h$ at the
Tevatron Collider. At hadronic colliders, in general, the possible
production processes for the Higgs are
\subequations
\label{eq:prod}
\bea
gg &\to& h
\label{proda} \\
q \bar q' &\to & h W 
\label{prodb}\\
q \bar q & \to & h Z 
\label{prodc} \\
q q &\to& h q q 
\label{prodd}\\
gg,q \bar q &\to& h t \bar t, h b \bar b.
\label{prode}
\eea
\endsubequations
At Tevatron energies, the most efficient processses are the first two 
of Eqs.~\ref{eq:prod}. The $W/Z$ produced in association with the $h$,
in Eqs.~\ref{prodb},\ref{prodc}, increases the viability of the signal in
the $b \bar b$ channel.  With the use of the high luminosity and a projected 
improvement in the `b' detection, it seems quite likely that the Tevatron
might be able to provide a glimpse of the light SM Higgs in the intermediate
mass range (IMR). The two new developements here have been the
use of $W W^*$ channel and/or the use of extra $b's$ in the
final state~\cite{han,sreerup} and use of neural networks~\cite{bhat} 
to increase the efficiency of the $b \bar b$ channel. Assuming that
it will be possible to achieve a $10\%$ resolution  for the  $b \bar b$ mass
reconstruction, an integrated luminosity of $\sim 30 {\rm fb}^{-1}$ is
required for  $3 \sigma - 5 \sigma $ signal for $\mh \sim 120$ GeV, when 
information from all the channels where $h$ is produced in association with
a $W/Z$ is combined. For the case of the MSSM Higgs a comprehensive analysis
has been done~\cite{carena}.  

At the LHC the way out is essentially to use the $h \to \gamgam$ channel
a B.R. which is about 1000 times smaller than that for the $b \bar b$ 
channel but  a considerably reduced background. To detect the Higgs in the
IMR in this channel the resolution required for $M_{\gamgam}$ measurement
is $\lsim 1 $ GeV $\simeq 0.1 \mh$~\cite{21}. 

The new developements  on the theoretical side have been new calculations
of various production cross-sections~\cite{spira} including the higher order 
corrections. On the analysis side the new developements have been the detector 
simulations for the \gamgam\ mode as well as the $b \bar b$ mode.
%\begin{figure}[htb]
%\begin{center}
%\vspace{-0.8cm}
%\mbox{\epsfig{file=../hadr13/lowmshig.eps,height=60mm}}
%\end{center}
%\vspace{-.3cm}
%\caption{The expected significance level of the SM Higgs signal at LHC for
%the intermediate mass region\label{fig3}}
%\end{figure}
\begin{figure}[htb]                
        \centerline{
            \psfig{figure=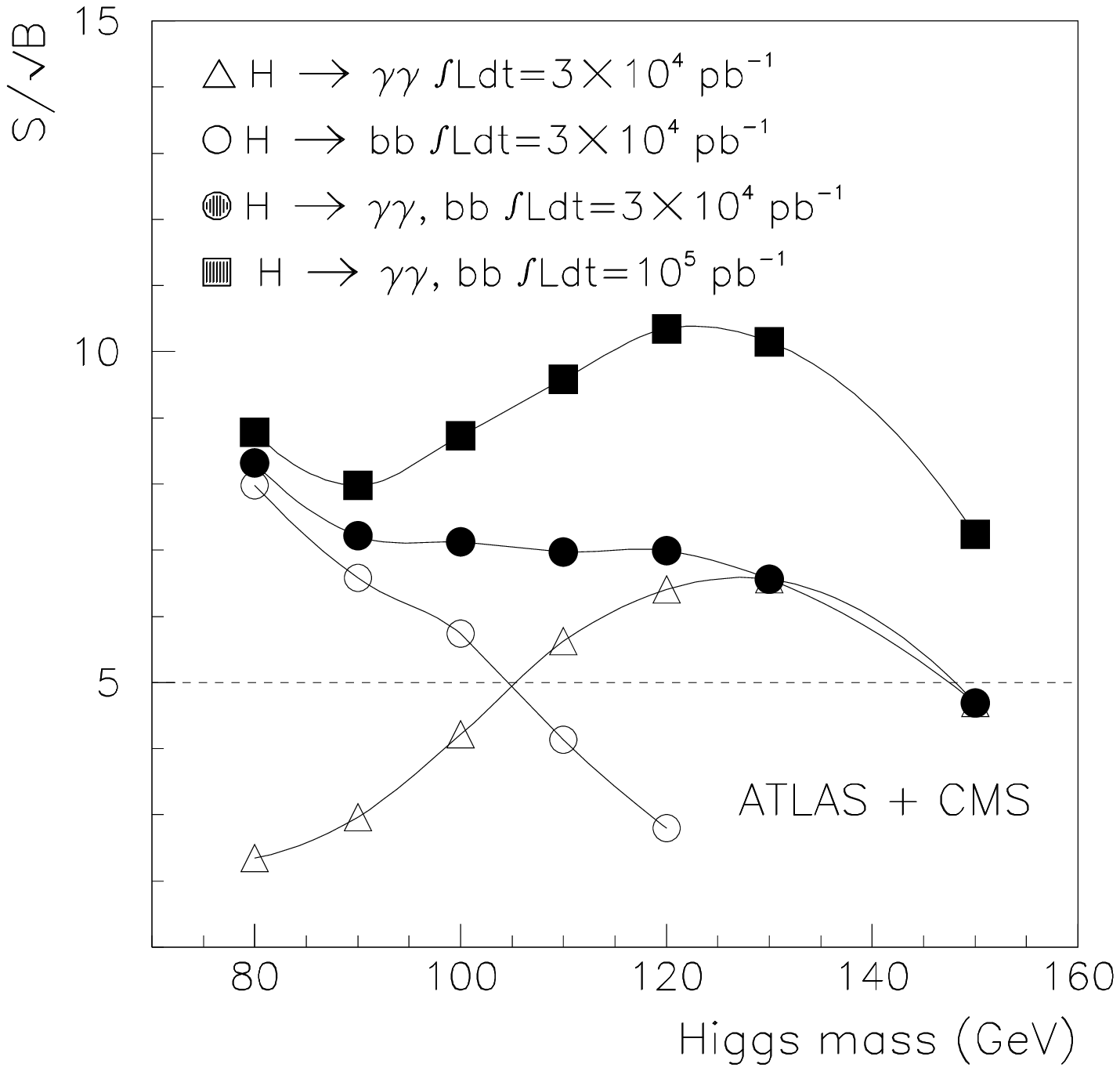,width=7cm,height=8cm,angle=0}
            \psfig{figure=chapter19_fig1.epsi,width=7cm,height=6.8cm,angle=0}
                   }
        \caption{\em The expected significance level of the SM Higgs
signal at LHC. The figure at left is from Ref.~\protect\cite{cms1}
and the one at right is taken from Ref.~\protect\cite{21}.
			     }
        \label{fig3}           
\end{figure}
The left panel in Fig.~\ref{fig3}, taken from the CMS/ATLAS techincal 
proposal~\cite{cms1}
shows the  expected $S/\sqrt{B}$ for the SM Higgs in the intermediate mass
range, using the $\gamgam , b \bar b$ modes. The use of $b \bar b$ mode 
for $\mh < 100 $ GeV is essentially achieved by using the associated 
production of Eq.~\ref{prodb},\ref{prodc}. Note that this will require an 
integrated luminosity of about $30~{\rm fb}^{-1}$, which is three years of 
LHC running at low luminosity.  The figure at the right shows the latest 
analysis of the achievable significance in the ATLAS detector over the
entire mass range, taken from Ref.~\cite{21}. This shows clearly how different
mass ranges are covered by different decay modes of the SM Higgs $h$.

A recent developement has been a 
demonstration~\cite{dieter} at the parton level, that the production of
the $h$  through the $WW/ZZ$ boson fusion processes of Eq.~\ref{prodd}
can be used for Higgs detection using the $\gamgam, \tau^+\tau^-$ and 
$WW \to l^+ \nu l^- \nu$ decay modes of the $h$. This has been
studied in the context of the ATLAS detector at LHC~\cite{les_houches}
with a point of view of exploiting this for measuring the relative 
ratios of the different Higgs couplings at the LHC.
 
For  the mass range (ii) the channels with the higher branching ratio, 
containing the $VV/VV^* (V=W/Z)$ are also the cleanest channels. The figure of
merit for a particular channel is clearly the value of the $\sigma 
\times {\rm B.R.} $. 
\begin{figure}[htb]
\begin{center}
\vspace{-0.8cm}
\mbox{\epsfig{file=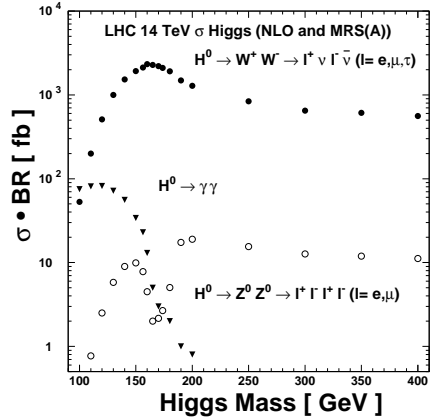,height=60mm}}
\end{center}
\caption{Expected $\sigma \times BR$ for different detectable SM Higgs
decay modes~\protect\cite{mdittmar}.\label{fig4}}
\end{figure}
Figure~\ref{fig4} taken from Ref.~\cite{mdittmar} shows this for the LHC
for both the mass ranges (i) and (ii).
The largest contribution to the cross-section here comes from the $gg$ fusion
(cf. Eq. ~\ref{proda}). Upto $\mh = 500 $ GeV the `gold plated signal'
with four charged leptons arising from the $h \to ZZ \to l^+ l^- l^+ l^-$ 
is the cleanest one and has been studied in great detail. In this channel
(and of course in the \gamgam\ channel) the Higgs can be reconstructed as
a narrow mass peak. ATLAS collaboration has demonstrated that for 
$\mh = 400 $ GeV, with a luminosity of only $10\; {\rm fb}^{-1}$ 
(one year of the low luminosity option of LHC),  one will see 27 signal events
as opposed to a background of $\simeq 10$ events.

The four lepton signature, however, suffers from low branching ratios for
$2 m_W  \lsim \mh < 2 m_Z$ and requires $30-100~{\rm fb}^{-1}$. In this
mass range the channel $h \to WW(W^*) \to l^+ l^- \bar \nu \nu$ 
provides a statistically significant signal with a luminosity
of $1-2\;{\rm fb }^{-1}$~\cite{mdittmar}.  The cross-sections 
for the associated process of Eqs.~\ref{prodb} \ref{prodc}, are still 
significant.  This can prove useful to see the Higgs in more than 
one channels (cf. the strategy adopted at the Tevatron). 

For $\mh > 500 $ GeV, detection of the Higgs as a narrow mass peak is no longer
feasible and the size of the four lepton signal is also very small. The best
chance for the higher mass Higgs is the detection, by using its production 
along with forward,high rapidity jets via the process of Eq.~\ref{prodd}.
For this higher mass range, the detection seems to
be a certainty at LHC, upto $\mh \sim 700$ GeV.
The processes of Eq.~\ref{prode} can, in principle, be used to determine the 
couplings of Higgs to heavier quarks.

\subsubsection{MSSM Higgs}
The Higgs sector is the one sector of the Supersymmetric theories where
some discussion can be carried out in a model independent way. For example,
upper bound on the mass of \hzer\ is quite robust  whether we consider 
MSSM or some extensions of it {\it i.e.} the (N)MSSM. The general qualitative 
observations about the couplings and the mass heirarchy among various scalars 
in the theory are also model independent. However, the different production
cross-sections and the decays do depend crucially on the superparticle spectrum.
Hence, while discussing the reach of future colliders, one discusses the 
SUSY Higgs search in the context of MSSM  with certain assumptions about the
particle spectrum. For large (small) values of \ma\ the \hzer (\Hzer) has mass
and couplings similar to the SM Higgs $h$.  For MSSM Higgs the discussion of 
the actual search possibilities is much more involved.  For the lightest 
scalar \hzer\  in the  MSSM, the general discussions of the intermediate mass 
SM Higgs apply, with the proviso that the \gamgam\ branching ratios are smaller 
for \hzer\ and hence the search that much more difficult.

In discussing the search strategies and propspects of the MSSM scalars one has 
to remember the following important facts:
\begin{enumerate}
\item Due to the reduction of the $h_0 W W$ coupling, 
the $h_0 \gamma \gamma$ coupling is suppressed as compared 
to the corresponding SM case. Of course one also has to include the
contribution of the charged sparticles in the loop~\cite{abdel2}.
The upper limit on \mhzer\ implies that  the 
decay mode into $WW (VV)$ pair is not possible for $h_0$, due to kinematic
reasons. On the other hand, for $H_0$ the  suppression of the coupling to $VV$ 
makes the decay less probable as compared to the SM case. 
As a result, the MSSM scalars are expected to be much narrower
resonances as compared to the SM case. For example, the  maximum width of 
$h_0$ is less than few MeV, for reasonable values of $\tan \beta$ 
and even for the heavier scalars $H_0$ and $A$,  the width is not more than few tens
of GeV even for masses as high as 500 GeV. 

\item $h_0$ is much narrower than the SM Higgs.
However, over a wide range of values of $\tan \beta$ and $m_A$, the $h_0$
has dominant decay modes into Supersymmetric particles. The most interesting
ones are those involving the lightest neutralinos, which will  essentially
give `invisible' decay modes to the $h_0, H_0$ and $A$ ~\cite{abdel1}.

\item On the whole for the MSSM scalars the decay modes into 
fermion-antifermion pair are the dominant ones due to the point 
(1) above as well as the fact that the CP odd scalar A does not have 
any tree level couplings to $VV$. Hence,
looking for the $\tau^+ \tau^-$ and $\bar b b$ final state becomes very
important for the  search of the MSSM scalars.
\end{enumerate}
As far as the lightest CP even neutral Higgs \hzer\  is
concerned, the major effects on the search prospects are three : 
\begin{enumerate}
\item  
The change in $gg \to  \hzer$  production cross-section due to the light stop
loops. These effects are sensitive to mixing in the stop-sector and so is
the mass of \hzer, the latter through radiative corrections.
\item 
A change in the $\hzer \to  \gamma \gamma$ width due to light sparticles
(specifically stops and charginos) in the loop.
\item Invisible decays of 
$\hzer \to  \tilde \chi_i^0 \tilde \chi_j^0$ (i,j = 1-4).
\end{enumerate}
Many of these have been  subject of detailed  investigations of 
late~\cite{abdel1,abdel2,fawzisridhar,fawzime}. 
The sizes of all these effects do depend on
the model parameters. All the three effects can conspire together to make
the \hzer\ 'invisible'.   Since both  the production and decay
are affected by supersymmetric effects, the information is best represented in
terms of 
\be
R_{gg\gamma\gamma} = 
{{\Gamma^{SUSY}(h\to gg) \times BR^{SUSY}(h \to \gamma \gamma)}\over
{\Gamma^{SM}(h\to gg) \times BR^{SM}(h \to \gamma \gamma)}},
\label{eq3}
\ee
and
\be
\label{eq4}
R_{\gamma \gamma} = {{{BR^{SUSY} (h \to \gamma \gamma) } \over
                    {BR^{SM} (h \to \gamma \gamma)}}}.
\ee
\begin{figure}[htb]
\begin{center}
%\mbox{\epsfig{file=../../daefigs/sf03xv.ps,height=80mm}}
\mbox{\epsfig{file=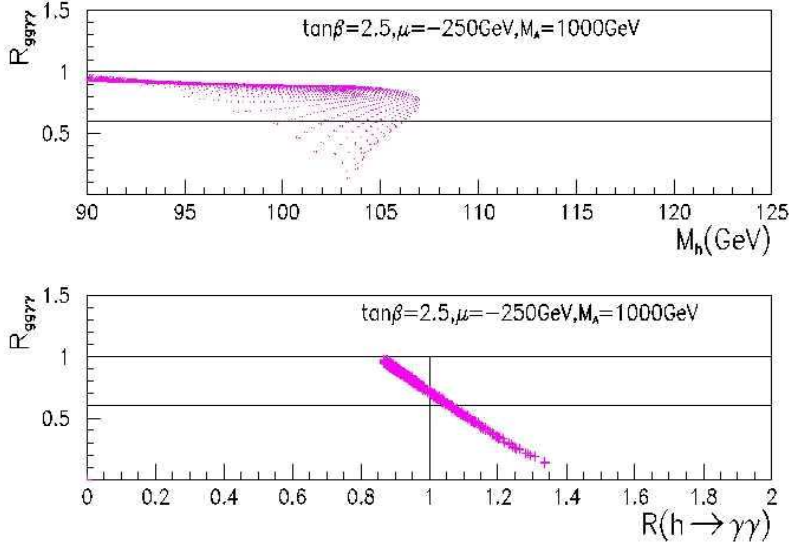,height=80mm}}
\end{center}
\caption{Ratio $R_{gg\gamma\gamma}$ of eq.~\protect\ref{eq3} as a function
of $R_{\gamma \gamma}$  of \protect\ref{eq4} and \mh\  
\protect\cite{fawzisridhar}. The values of various 
parameters are indicated in the figure. \label{fig5}}
\end{figure}
Fig.~\ref{fig5} reproduced from Ref. \cite{fawzisridhar} shows the ratio
$R_{gg\gamma\gamma}$ as a function of 
\mh\ and $R_{\gamma \gamma }$ for choices of parameters mentioned in the figure.
The  depletion in the ratio $R_{gg\gamma \gamma} $  here is mainly 
due to the small $m_{\tilde {t_1}}$.
The investigations try to focus on the fact that eventhough the inclusive
$ 2 \gamma$ signature is substantially reduced the associated production 
via processes of the Eqs.~\ref{prodb}, \ref{prodc} and \ref{prode} can 
still provide a viable discovery channel for the light, MSSM Higgs. 
Also production of Higgs in decays of stops (which are light and
hence have large production cross-sections) provides an 
additional channel.   

Luckily the light charginos and
neutralinos affect the inclusive $2 \gamma$ channel only in small regions 
of parameter space~\cite{fawzime}, once LEP constraints are imposed on the
chargino/neutralino sector. The decays of $h$ into invisibles can still
contribute to the problem though. Under the assumption of a common gaugino
mass at high scale, a dangerous reduction in BR $ (\hzer \to  \gamgam)$ is
possible only in the pathological case of a degenerate sneutrino and
chargino. However, for nonuniversal gaugino masses which predict light
neutralinos, even after eliminating the region which would give too large
a relic cosmological density, there exist regions in parameter space where
the usual $\gamma\gamma$ signal for \hzer\ drops drastically. 
\begin{figure}[htb]
\begin{center}
%\mbox{\epsfig{file=../../daefigs/rf01xv.ps,height=60mm}}
\mbox{\epsfig{file=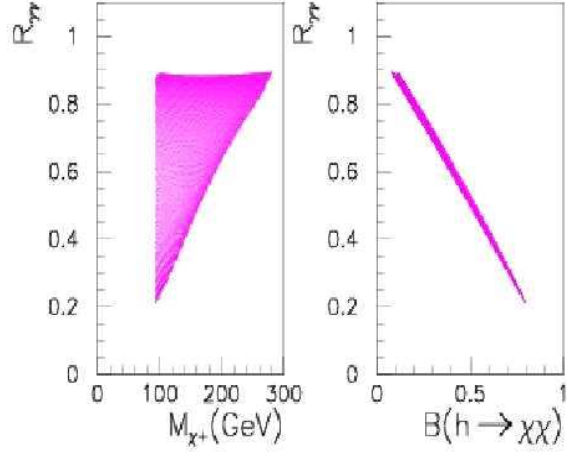,height=70mm}}
\end{center}
\caption{Ratio $R_{\gamma \gamma}$  of eq. \protect\ref{eq4}  as a  
function of $m_\chi^+$ and the `invisible' decay width of the $h$
for nonuniversal gaugino masses with $M_2= 10 M_1$, for heavy 
selectrons~\protect\cite{fawzime}. 
\label{fig6}}
\end{figure}
Fig~\ref{fig6} shows, again for the values of parameters mentioned in the 
figure caption, $R_{\gamma \gamma}$ as a function of $M_{\chi^+}$ and 
$B(h \to \tilde \chi^0 \tilde \chi^0)$. In this case, unlike the case 
of the light stops, the production of Higgs in decays of charginos/neutralinos
does not have very high rates and the search will suffer from the same 
reduction of the $\gamma \gamma$ channel due to decay into invisibles. Thus 
in  this scenario, for large values of the mass of the CP odd Higgs, 
a signal for SUSY through the Higgs sector may
not be feasible through direct search in the $\gamma \gamma$ channel. 

Thus we see that the detection of the lightest SUSY  Higgs at
Tevatron/LHC will be difficult, but feasible. It will surely require 
high luminosity run. Recall again here that the low value of \mh\
is the preferred one by the EW measurements~\cite{19} and also expected if
weak scale SUSY is a reality.

However, since in the MSSM there exist many more scalars in the spectrum 
one can cover the different regions in the parameter space by looking also 
for $A,\Hzer$ and $H^\pm$. 
\begin{figure}[htb]
\begin{center}
%\mbox{\epsfig{file=../../daefigs/chapter19_fig4.epsi,height=60mm}}
\mbox{\epsfig{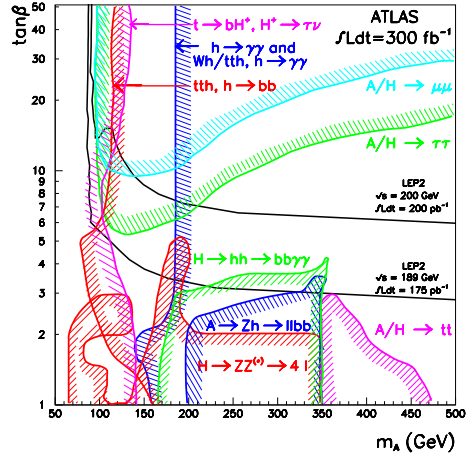}}
\end{center}
\caption{Five $\sigma$ discovery contours for the MSSM Higgs for
the ATLAS detector with 300 ${\mathrm fb}^{-1}$ 
luminosities~\protect\cite{21}.\label{fig7}}
\end{figure}
Fig.~\ref{fig7}  taken from Ref.~\cite {21} shows the contours for 
5 $\sigma$ discovery level for different scalars in the MSSM in the plane of 
two parameters $\tan \beta$ and \ma\ . 
Thus we see that at low values of \ma, almost all the  scalars  of the MSSM 
are kinematically acessible at LHC.  However,  at large \ma\ ,
which seem to be preferred by the data on $b \to s \gamma$ and also by the
ever upward creeping lower limit on \mh\ from direct searches, 
even after combining the information from various colliders 
(LEP-II, Tevatron (for the charged Higgs search) and of course LHC), 
a certain region in the $\ma - \tan \beta$ plane remains inaccessible. 
This hole can be filled up only after combining the data from the CMS and 
ATLAS detector for 3 years of high luminosity run  of LHC.  Even in this 
case there exist regions where one will see only the single light scalar.

\subsection{Establishing the quantum numbers of the Scalar}
Thus we see that the LHC can see at least one scalar, no matter what
its mass. However, to establish such a scalar as {\it the} Higgs, 
one needs to establish  two things, {\it viz.}
\begin{enumerate}
\item The scalar is CP even and has $J^P = 0^+$,
\item The couplings of the scalar with the fermions and gauge bosons
are proportional to their masses.
\end{enumerate}
This is also essential from the point of view of being able to distinguish 
this scalar from the lightest scalar expected in the MSSM.   In general the
coplings of the lightest scalar  \hzer\ are different from the SM Higgs
$h$. However, it should be kept in mind that in the large \ma\ region where 
the mass bound for \hzer\ is saturated, these couplings differ very little
for the two.  As a matter of fact this issue has been a subject of much 
investigation of late~\cite{tevwg,gunion}. The Snowmass Studies~\cite{tevwg} 
indicate that for a light Higgs ($\mh\ = \mz$)  such a discrimination is 
possible only to an accuracy of about 30 \%. The idea of using  the 
$h/\hzer$  production via $WW/ZZ$ fusion, to determine the ratios of 
couplings of the Higgs to different particles, is being studied 
now~\cite{les_houches}.
There are also interesting investigations~\cite{gunion} which try to device 
methods to determine the CP character of the scalar using hadron colliders.
It is in these two respects that the planned \eplem\ colliders~\cite{6}
can be a lot of help.

\subsection{Search of Higgses at $e^+e^-$ colliders:} 
Eventhough we are not sure at present 
whether such colliders will become a reality,  the technical 
feasibility of buliding a $500$ GeV \eplem\ (and perhaps an  attendant
\gamgam, $e^-e^-$ collider) and doing physics with it is now 
demonstrated ~\cite{6}.
We will see below that such a collider  can play a complementary  role and 
help establish the quantum numbers of the scalar mentioned above.
At these colliders, the production processes are $\eplem \to Z(^*) h$,
$\eplem \to e^+ e^- h, \eplem \to \nu \bar \nu h$ ,$\eplem \to t \bar t h$
and similarly  the associated production of h with a pair 
of stops $\tilde t_1 \tilde t_1 h$.

Detection of the Higgs at these machines 
is very simple if the production is kinematically allowed, 
as the discovery will be signalled by some very striking features of the 
kinematic distributions.  Determination of the spin of the  produced 
particle in this case  will also  be simple as the expected angular
distributions will be very different for scalars with even and odd parity. 
For an \eplem\ collider with $\rts \le 500$ GeV, more than one 
of the MSSM Higgs scalar will be visible over most of the parameter
space~\cite{6,spira,abdel3}.
Even with this machine one will need a total luminosity of 
$200\;\; {\mathrm fb}^{-1}$, to be able to determine the ratio of 
$BR (h \to c \bar c) / BR(h \to b \bar b)$, to about 
$7 \%$ accuracy~\cite{gunion}. The simplest way to determine 
the CP character of the scalar will be to produce $h$ in a \gamgam\ 
collider, the ideas for which are under discussion.

At large \ma\ (which seem to be the values preferred
by the current data on $b \to s \gamma$), the SM Higgs and \hzer\ are 
indistinguishable as far as their couplings are concerned. 
A recent study gives the contours of constant values for the ratio 
$$
\frac{BR(c \bar c)/ BR(b \bar b)|_{\hzer}} {BR(c \bar c)/ BR(b \bar b)|_{h}},
$$
as well as a similar ratio for the $WW^*$ and $b \bar b$ widths as a function
of $\tan \beta$ and \ma . 
\begin{figure}[htb]
\begin{center}
\vspace{-0.8cm}
%\mbox{\epsfig{file=../hadr13/test_gunion.eps,height=70mm}}
\psfig{file=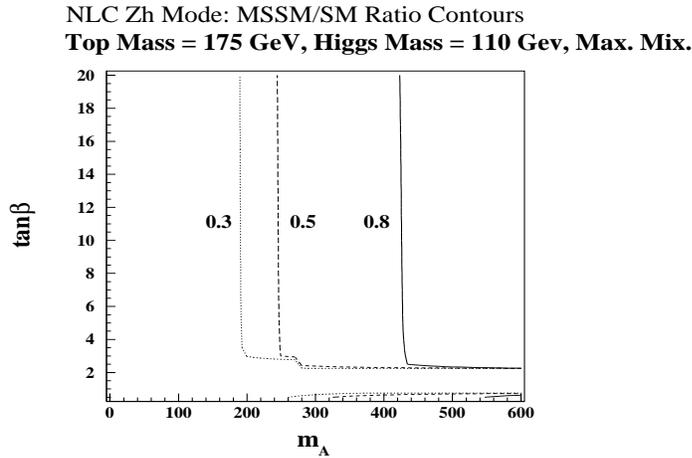,height=70mm,width=100mm}
\end{center}
\caption{The ratios of relative branching fractions for the MSSM and SM 
for maximal mixing in squark sector. The specific value of the Higgs mass
used is theoretically disallowed at large \ma\ and around $\tan \beta \sim 2$
~\protect\cite{tevwg}.\label{fig8}}
\vspace{-0.5cm}
\end{figure}
As we can see from Fig.~\ref{fig8}~\cite{tevwg}, a measurement of this 
ratio to an accuracy of about $10 \%$ will allow distinction 
between the SM Higgs $h$ and MSSM Higgs \hzer\ upto about 
$m_A = 500 $ GeV. As stated above, NLC should therefore
be able to do such a job.
Certainly,  the issue of being able to determine the quantum numbers and 
the various couplings  of the scalar including the self 
couplings  forms the subject of a large number of
investigations currently~\cite{6,spira,abdel4}.

\section{Supersymmetric particles other than Higgs}
 As explained in the introduction, Supersymmetry is the only theoretical
framework which can give stabilization of the weak scale against radiative
corrections and which has very definite predictions for the presence
of additional particles. The search strategies for these sparticles are kind 
of prototypes for all the searches for physics beyond the SM and are used 
to define the detector requirements. SUSY has model independent predictions 
about the spin and electroweak/strong couplings of these sparticles. One
has to take recourse to a specific model when we talk about their
mass spectrum. They are broadly related to the patterns of SUSY breaking.
The ones most often discussed are
\begin{enumerate}
\item{Gravity Mediated SUSY breaking},
\item{Gauge Mediated SUSY breaking (GMSB)},
\item{Anomaly Mediated SUSY breaking}.
\end{enumerate}
The expected mass patterns and the candidate for the lightest
supersymmetric particle (LSP) are different in each case.
In the first case the lightest supersymmetric particle is a neutralino
$\tilde \chi^0_1$  which is a mixture of Higgsinos and the Electroweak gauginos. In
the case of GMSB models the LSP is a 'light' gravitino and usually the
next lightest supersymmetric particle also behaves like a LSP. In this case,
production and decay of the sparticles at colliders produce final states
with photons, whereas in the former the final state has $\tilde \chi^0_1$ which
may/may not be stable depending on whether the $R$-$parity$ is conserved or
violated. In the former case the final state will contain a large amount of 
missing energy and in the latter case a large number of quarks/leptons.
In general the discussion of search for sparticles at the
current and future colliders has to cover all these possibilities. In the
case of gravity mediated SUSY breaking again there are options of
considering the constrained, predictive SUGRA framework (where the number
of additional parameters of the MSSM goes down from 124 to 5 due to various
assumptions) or looking at some model independent aspects. The decay patterns
of various sparticles depend crucially on the mass patterns and hence on the
assumptions one makes. For example, even in the option of gravity mediated
SUSY breaking, there exist virtual LSP's  along with $\tilde \chi^0_1$ 
in some regions of parameter space.  Such virtual LSP's can change the 
phenomenology of the sparticle  searches substantially~\cite {AD1,ambrasanio}. 
A large fraction  of the simulation studies done~\cite{7P,6P} so far 
have been in the context of  (M)SUGRA, with a few discussions~\cite{mura}
of the effect of relaxing the assumption of common gaugino mass at the 
high scale ~\cite{krasnikov} and that of common scalar mass 
at high scale ~\cite{AD,AD2} or both and more~\cite{Xrecent},
having started relatively recently.

Before we begin discussions about search strategies for the sparticles,
let us note that, the direct searches for sparticles at different \eplem\ 
and $p  \bar p$  colliders have so far come up with null results. The only 
hint of 
the existence of the sparticles is in the unification of the SU(2), U(1) and 
SU(3) couplings which  happens only in  SUSY-GUTS.  In this talk I will 
restrict myself to sparticle searches only in the scenario 1 of SUSY 
breaking from the above list  with and without \rpv\ conservation. As already
mentioned before, currently the focus of various phenomenological 
investigations is not so much on the search strategies for sparticles but on
the study of how well the soft Supersymmetry breaking parameters can be 
determined once we find the evidence for sparticles. As can be seen 
from various studies~\cite{cms1,21,6,7P,6P,8P} the TeV colliders, {\it viz.} 
Tevatron Run II/ Run III, LHC as well as the NLC should all be able to 
see the signal for the production and decay of sparticles if 
the weak scale SUSY is a reality.

\subsection{Discovery Potential for SUSY at the different colliders}
     At the hadronic colliders the sparticles with largest production cross-
sections and hence highest discovery potential, are $\tilde g,\tilde  q $. 
The $\tilde g, \tilde q$ are produced via 
$ gg, q \bar q \to  \tilde g \tilde g$ 
and  $gg, q \bar q \to  \tilde q  {\bar{\tilde q}}$.
The possible decay modes and various branching ratios clearly depend on
the mass spectra. Possible decay modes relevant for the LHC range are 
$  \tilde q \to  q \tilde \chi^0_j$  and 
$\tilde g \to  q \bar q \tilde \chi_j^0,
\tilde g \to q \bar q' \tilde \chi^{\pm}_i,$ with $j=1,2,3,4$ and $i=1,2$
depending upon the masses. These will then be followed by further 
decays of the charginos and heavier neutralinos ending 
in a $\tilde \chi^0_1$ which is stable for 
$R$-$parity$  conservation case and will give rise to missing
energy in the event. For \rpv\  one gets large number of leptons/
quarks in unusual combinations due to $\tilde \chi_1^0$ decay, 
in addition to the other particles in the former case. The cascade decays 
can give a very characteristic signal with real $Z/W$ s in the final states 
if kinematically allowed. Due to the rising importance of cascade decays 
for the larger masses of $\tilde g, \tilde q$, a good signal for 
$\tilde g, \tilde q$ production and decay is a final
state with m jets $(m>0)$, n leptons $(n\ge 0)$ and large missing 
transeverse energy. The Majorana nature of gluinos can give rise to 
like-sign-dilepton events. So the expected events are $0l (1l)$: 
Jets, $E_T^{\rm miss}$  and no(1) leptons, SS: same sign dileptons, OSS: 
Opposite side dileptons and $3l$: trileptons. For the most commonly 
expected sparticle mass spectra, the lighter chargions/neutralinos are 
among the lightest sparticles. 
These will give rise to the very interesting final state containing 
only  leptons and missing energy via (e.g.)
$p \bar p(p)  \to  \tilde \chi^\pm_1 + \tilde \chi^0_2 $
$\to l^\pm\tilde \chi_1^0  Z^* \tilde \chi_1^0$. 
These 'hadronically quiet' trileptons are a very clean channel for SUSY
discovery.  The assoicated production of the
gauginos is thus signalled by $3l,0j$: trileptons with a jet veto or 
$2l,0j$: dileptons with a jet veto. These have been used for SUSY search 
even at the current Tevatron studies~\cite{xer-tri}.  Higher order corrections 
to the production cross-sections of the gluinos/squarks and gauginos
are now available~\cite{spira_les}.  The discovery potential of LHC 
for leptonic channels in the constrained (M)SUGRA scenario is shown in 
\begin{figure}[htb]
\begin{center}
%\mbox{\epsfig{file=../../daefigs/figrea.epsi,height=60mm}}
\mbox{\epsfig{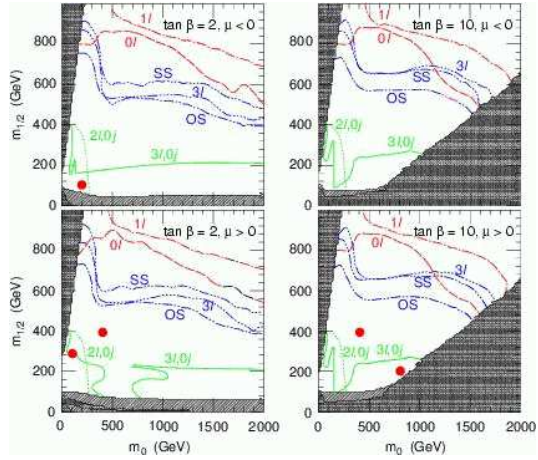}}
\end{center}
%\vspace{-.3cm}
\caption{Reach for $S/\sqrt{B} > 5$ for various SUSY signatures in SUGRA
parameter space. Various symbols are explained in the text. Shaded  regions
are disallowed either by current searches or theoretical 
considerations~\protect\cite{21}.\label{fig9}}
\end{figure}
Fig.~\ref{fig9}. The effect of, eg. nonuniversality in 
scalar/gaugino masses at high scale, on these analyses can be substantial 
and is just beginning to be explored.  One of the squarks of the third 
generation viz. $\tilde t_1$ can be substantially
lighter than all the others. As a result, search strategies for the light
stop are entirely different~\cite{lighttopus,7P,6P}. As already mentioned in
the discussion of the MSSM Higgs associated production of $\tilde t_1
\tilde t_1 \hzer$ can be an interesting discovery channel at the 
LHC/NLC~\cite{fawzisridhar,abdel6}. 

\subsection{Determination of the soft Supersymmetry breaking parameters}
Since production of different sparticles can give rise to the same final 
states, the real problem at a hadronic collider  will be to 
seperate signals due to different sparticles. The observed signal 
distributions are sums of products of production cross-sections, 
branching ratios and acceptances. Hence it seems that a model independent 
interpretation is impossible. Luckily some kinematical quantities can be 
extracted in model independent ways using some characteristic decay 
distributions. Events near the end point of the 
$m_{ll}$ distribution for three body decay of $\tilde \chi_2^0 \to 
\tilde \chi^0_1 l^+ l^-$ (decay caused either by a virtual Z or $\bar l$) 
play a very important role in reconstructing the kinematics of 
$\tilde g/\tilde q$ cascade decay chain. This can then
be used to reconstruct the (M)SUGRA parameters~\cite{hinchliff}. It is
important to investigate model independence of such reconstructions. A study
~\cite{Mihoko1,Mihoko2,Mihoko3} shows that the resolution of 
the $m_{ll}$ distribution
end point depends on the square of the matrix element for the 
decay and can introduce additional systematic errors in the extraction of 
SUSY model parameters from kinematics. On the positive side, the decay 
distributions can give nontrivial information on slepton masses and 
mixing ~\cite{Mihoko3,Iashvelli}.

      The TeV energy e+e- colliders in planning ~\cite{6} will play a 
very useful and complementary role. Of course, only particles with 
electroweak couplings viz.  squarks, sleptons, charginos/neutralinos and 
Higgses can be produced at these machines. Quite a few detailed studies 
of the production of the sfermions and chargino/neutralinos exist in 
literature in the context of $R$-$parity$ conserving 
SUSY~\cite{XT1,XT2,XT3,MP,FM,martyn} as well as in the context of \rpv\
SUSY ~\cite{ghoshme}. 
\begin{figure}[htb]
\vspace{-0.5cm}
\begin{center}
%\mbox{\epsfig{file=../rpv_nlc/Fig4.ps,height=80mm}}
\mbox{\epsfig{file=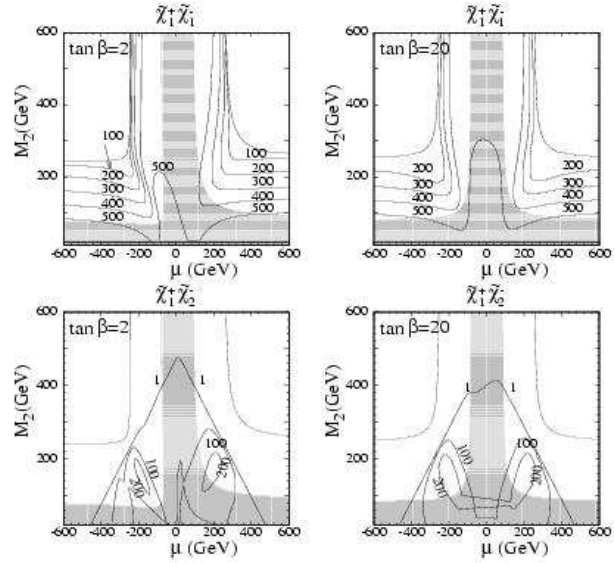,height=80mm}}
\end{center}
%\vspace{1.5cm}
\caption{Chargino production cross-sections at an \eplem\ 
collider  with $\sqrt{s} = 500$ GeV, with the assumption of
a universal gaugino and scalar mass at high scale, with 
$M_{{\tilde e}_L}= M_{{\tilde e}_R} = 150$ GeV~\protect\cite{ghoshme} 
\label{fig10}}
\end{figure}
The cross-sections for the sfermion productions are given completely
in terms of their EW quantum numbers and masses, except for those of
the third generation where they also depend on the $L-R$ mixing in the
sfermion sector induced by the soft supersymmetry breaking terms.
For charginos/neutralinos the cross-sections depend on the SUSY
parameters in a nontrivial way.
Fig.~\ref{fig10} shows contours of constant cross-sections (in fb)
for the production of a pair of charginos. The shaded area is 
ruled out by LEP constraints  and the dotted lines show
the kinematical limit for chargino production. Thus one sees that 
the production cross-sections are quite large.
In the case with $R_p$ conservation, a systematic study of the
possible accuracy of the kinematical reconstruction of various 
sparticle masses and a test of different (M)SUGRA mass relations 
has been performed. 
\begin{figure}[htb]
\begin{center}
%\mbox{\epsfig{file=../../daefigs/mur_bb/smu-mass.eps,height=50mm}}
\mbox{\epsfig{file=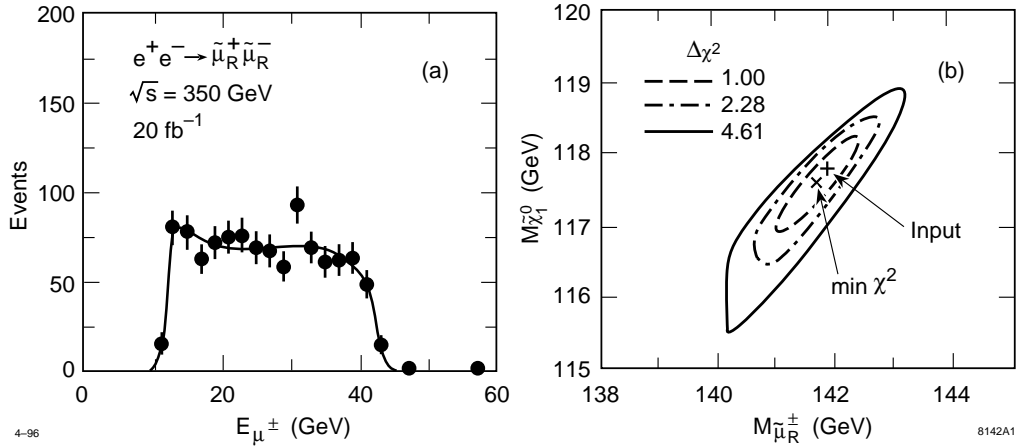,height=60mm}}
\end{center}
\caption{An example of  the possible accuracy of the determination 
of the smuon mass and neutralino mass, for a $e^+e^-$ collider with
energy and lumionisty as mentioned in the figure~\protect\cite{XT2}.
\label{fig11}}
\end{figure}
Fig.~\ref{fig11} taken from Ref.~\cite{XT2} shows the accuracy of the possible 
reconstruction of the masses of the smuon and  the lightest neutralino  using 
kinematic distributions using the slepton production. Using this along with  
the absolute value of the cross-sections with polarised $e^+/e^-$ beams
and angular distributions of the produced sleptons, one can
extract the SUSY breaking parameters $M_2,\mu$ and $\tan \beta$. 
It should be then possible to test 
1)the (un)equality of (e.g.) $M_{{\tilde e}_R}$ and $M_{{\tilde \mu}_R}$,~
2)as well as the assumption of a common scalar mass at high scale.
\begin{figure}[htb]
\begin{center}
%\mbox{\epsfig{file=../../daefigs/mur_bb/semasses.eps,height=50mm}}
\mbox{\epsfig{file=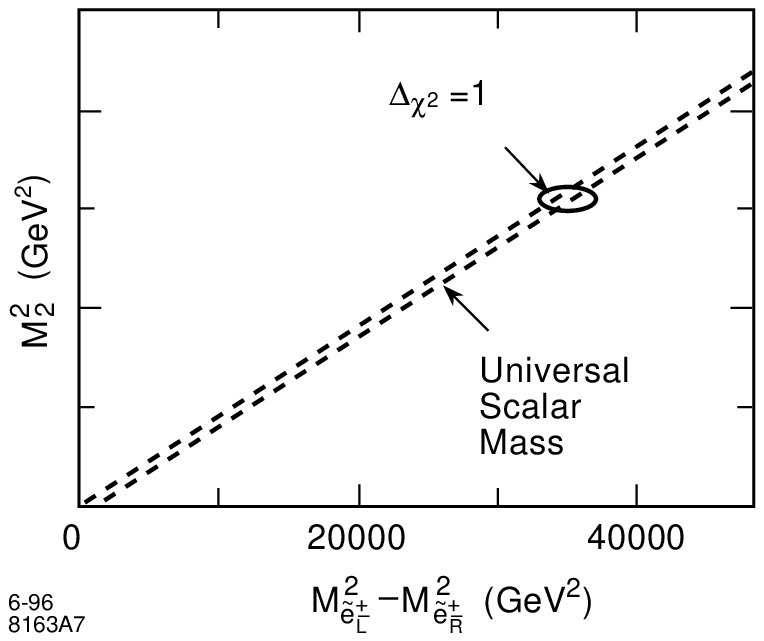,height=50mm}}
\end{center}
\caption{Possible accuracy of a test of universal scalar mass possible
at NLC with $\sqrt{s} = 350$ GeV and luminosity $20 {\rm fb}^{-1}$
from the measurements of masses of $\tilde e_R$ and 
$\tilde e_L$\protect\cite{MP}\label{fig12}}
\end{figure}
Fig.~\ref{fig12} shows the possible accuracy of reconstruction of $M_2^2$
and $M^2_{{\tilde e}_R}$ - $M^2_{{\tilde e}_L}$, which can thus test the
assumption of  universal scalar mass. It should be emphasized here that
the directions of search for SUSY at the NLC will be largely defined 
by what we find at the LHC. Relating the  measurements of $M_2,M_1$ etc. 
at the NLC with the results at the LHC, will then allow us to arrive at 
an understanding of the soft supersymmetry breaking parameters. 

It has been recently demonstrated~\cite{abdel5} how using polarised $e^+/e^-$ 
beams and studying angular distributions in chargino production one can 
reconstruct the parameters of the chargino sector. Issues under discussion now
are also effects of higher order QCD corrections ~\cite{man_eboli,sabine,dekg} 
on  the production and decay of squarks, on the accuracy of kinematic 
determination of squark mass ~\cite{feng_finnel} or possibility of 
using the highly precise measurements at NLC with polarization to test 
the equality of fermion-fermion-gauge boson and fermion-sfermion-gaugino 
couplings~\cite{FM,others}.

Thus in summary,  we see that  LHC will certainly be able to see 
evidence for sparticles.
Using clever use of kinematic distributions it seems possible to
reconstruct different soft SUSY breaking parameters from first 
recosntructing  various masses kinematically. Along with the NLC
one should be able to disentangle different contributions at a hadronic
collider from each other and determine the soft SUSY breaking parameters.
 
\section{Conclusions} 
\begin{enumerate}
\item  The current experimental information from LEP and LEP-II
as well as the direct measurement of the top mass indicates that
a light Higgs boson is likely. LHC should be able to see a Higgs boson
close to $2 M_W$ threshold  reasonably easily. Heavier ones are also easily 
detectable upto about 600 GeV. In the Intermediate Mass Range between the
LEP limit and $2M_W$ one would require the high luminosity, but the Higgs
signal would be clear.
\item With the current information and constraints from LEP as well 
as $b \to s \gamma$, the decoupling scenario  is becoming more and 
more likely for the SUSY Higgs. At least one Higgs (and in some region
of the parameter space  two Higgses) can be seen at LHC. However, there 
still exists a  region in the $\tan \beta - \mu$ plane, where the detection 
of Higgs signal is very difficult if not impossible.
\item Discrimination between a SM and a MSSM Higgs using only LHC seems
difficult.
\item LHC can see signals for all  non-Higgs  sparticles  if they 
exist at mass scales expected in the weak scale SUSY.
\item An \eplem\ collider with $ \rts \ge 350 $ GeV can effectively
see at least two of the five Higgses, if they are within kinematic reach,
independent of any other parameter. The parity of the scalar produced 
can be trivially determined at an \eplem\ collider. It seems quite difficult 
to determine the $CP$ character of the scalar produced  using only hadronic 
and \eplem\ colliders and $\gamma\gamma$ colliders might be needed for that.
At a $\mu^+ \mu^-$ collider, separation between  $h$ and $h_0$ based
on measurements of relative branching ratios is possible upto $\ma = 500$ 
GeV.
\item Using special kinematical features of the decay distributions,
it seems possible (though it needs much more study) to  determine some
of the soft SUSY breaking parmeters even at a hadronic collider. However, 
a TeV scale \eplem\  collider  along with LHC can indeed afford a very clear 
determination of the soft SUSY breaking parmeters if the sparticles 
are kinematically accessible. 
\end{enumerate}
\section*{Acknowledgements} I wish to thank the organisers of the XIII th
DAE symposium for organising the DAE symposium very efficiently. 
I wish to  acknowledge the Department of Science and Technology (India)
and the  National Science Foundation's U.S. India Cooperative Exchange 
Program for partial support under the NSF grant INT-9602567.


\begin {thebibliography}{99}
\bibitem {1} See for example, talk by A. Gurtu in these proceedings.
\bibitem{LEPHWG} LEP WG for Higgs Boson Searches, {\bf CERN-EP/99-060}, 1999.
\bibitem{Moriond} Aleph 2000-028, Delphi 2000-050, L3 Note 2525, OPAL
Technical Note TN646; Contributions of the LEP collaboration to 
`Recontres de Moriond', Les Arcs, France, March 11-25,2000.
\bibitem {2} See for example, talk by J. Maharana in these proceedings.
\bibitem {3} For one of the earliest discussions see, 
R. Kaul, \Journal{\PLB}{109}{19}{1982}.
\bibitem{cms1} CMS Technical proposal, {\bf CERN/LHCC/94-38} (1994); 
ATLAS technical proposal, {\bf CERN/LHCC/94-93} (1994).
\bibitem{21} ATLAS Technical Design Report 15, {\bf CERN/LHCC/99-15} (1999).
\bibitem {6} ECFA/DESY/LC physics Working Group ,
A. Accomando et al, \Journal{\PR}{299}{1}{1998}.
\bibitem {5} {\em Perspectives on Higgs Physics I and II}, Ed. G. Kane,
World Scientific, Singapore (1997).
\bibitem{20} For example see, Run II Higgs Working Group (Run II SUSY/Higgs
Workshop) http://fnth37.fnal.gov/higgs.html.
\bibitem{9P} The Higgs Working Group Report, Physics at TeV 
Colldiers (Les Houches), A. Djouadi et al, {\bf hep-ph/0002258}.
\bibitem {4} {\em Perspectives on Supersymmetry}, Ed.  G. Kane, World 
Scientific, Singapore (1998).
\bibitem{7P} SUGRA working Group report, Run II 
SUSY Working Group (Run II SUSY/Higgs Workshop) 
http://fnth37.fnal.gov/susy.html. 
\bibitem{6P}SUSY Working Group Report, Physics at TeV Colliders 
(Les Houches)~http://lappc-th8.in2p3.fr/Houches99/susygroup.html.
\bibitem{8P}M. Carenna, R. Culbertson, H. Frisch, S. Eno and S. Mrenna,
\Journal{\RMP}{71}{937}{1999}.
\bibitem {7} D. L. Burke et al., \Journal{\PRL}{79}{1626}{1997}.
\bibitem {8} R.M. Godbole, Proceedings of {\it Workshop on Quantum 
aspects of Beam Physics}, Jan. 4-Jan. 9, 1998,Monterey, U.S.A.,
pp. 404-415, Ed. P. Chen,  World Scientific (1999), {\bf hep-ph 9807379}.
\bibitem {9} P. Chen, T. Barklow and M.E. Peskin, 
\Journal{\PRD}{49}{3209}{1994}.
\bibitem {10} M. Sher, \Journal{\PR}{179}{273}{1989}.
\bibitem {11} M. Lindner, \Journal{\ZPC}{31}{295}{1986}.
\bibitem{12}G. Altarelli and G. Isidori, \Journal{\PLB}{337}{141}{1994};
J.A. Casas, J.R. Espinoza and M. Quiros, \Journal{\PLB}{342}{171}{1995};
\bibitem{13} T. Hambye and K. Riesselmann, \Journal{\PRD}{55}{7255}{1997};
{\bf hep-ph/9708416} in {\em ECFA/DESY study on particles and detectors for 
the linear colliders}, Ed. R. Settlers, {\bf DESY 97-123E}.
\bibitem{14} H.E. Haber and Y. Nir, \Journal{\NPB}{335}{363}{1990}.
\bibitem{15} A. Djouadi, J. Kalinowski and and P.M. Zerwas, 
\Journal{\ZPC}{57}{569}{1993}.
\bibitem{16} G.L. Kane, C. Kolda and J.D. Wells, \Journal{\PRL}{70}{2680}
{1993}
\bibitem{17} J.R. Espinoza and M. Quiros, \Journal{\PLB}{302}{51}{1993}.
\bibitem{18}P.N. Pandita, \Journal{\pramana}{51}{169}{1998} and 
references therein, in {\em Proceedings of the Workshop on High Energy
Physics Phenomenology, V, Pune, India}, Eds. R.V. Gavai and R.M. Godbole, 
(Indian Academy of Science, Bangalore, 1998).
\bibitem{heinemeyer} S. Heinemeyer, W. Hollik and G. Weiglein, 
\Journal{\PLB} {455}{179}{1999};\Journal{\EUJ} {9}{343}{1999}. 
\bibitem{abdel1} A. Djouadi, P. Janot, J. Kalinowski and P.M. Zerwas, 
\Journal{\PLB} {376}{220}{1996}.
\bibitem{abdel2} For a summary of the effects of sparticles see:
A. Djouadi, \Journal{\PLB}{435}{1998}{101}, ({\bf hep-ph 9806315})
and references therein.
\bibitem{spira} M. Spira and P.M. Zerwas, 
{\em Lectures at the Internationale Universit\"atwochen f\"ur Kern und
Teilchen Physik}, {\bf hep-ph/9803257} and references therein.
\bibitem {19} LEP Electroweak Working Group, {\bf CERN-EP-2000/016},2000.
\bibitem{han} T. Han and R.J. Zhang, \Journal{\PRL}{82}{25}{1999},
T. Han, A.S. Turcot and R.J. Zhang, \Journal{\PRD}{59}{093001}{1999}.
\bibitem{sreerup}
D. Choudhury,  A. Datta and S. Raychaudhury, {\bf hep-ph/9809552}.
\bibitem{bhat} P.C. Bhat, R. Gilmartin and H.B. Prosper,
{\em Fermilab-Pub-00/006} {\bf hep-ph/0001152 v2}.
\bibitem{carena} M. Carena, S. Mrenna and C.E.M. Wagner,
\Journal{\PRD}{60}{075010}{1999}{\bf hep-ph 9808312 v2};
H. Baer, B.W. Harris and X. Tata, \Journal{\PRD}{59}{015003}{1999}. 
\bibitem{dieter} D. Rainwater and D. Zeppenfeld, \Journal{\JHEP}{12}{5}{1999};
D. Rainwater, D. Zeeppenfeld and K. Hagiwara, \Journal{\PRD}{59}{14037}{1999};
T. Plehn, D. Rainwater and D. Zeppenfeld, {\bf hep-ph 9911385}; D. Rainwater
and D. Zeppenfeld, \Journal{\PRD}{60}{113004}{1999} erratum to appear ({\bf
hep-ph 9906218 v3}).
\bibitem{les_houches} D. Zeppenfeld, R. Kinnunen, A. Nikitenko and E. Richter-
W{\c{a}}s, {\bf hep-ph 0002036}.
\bibitem{mdittmar}M. Dittmar, {\em ETHZ-IPP} PR-98-10, ({\bf  hep-ex 9901009}).
\bibitem{fawzisridhar} G.Belanger, F. Boudjema and K. Sridhar,
\Journal{\NPB}{568}{3}{2000}~({\bf hep-ph 99004348}).
\bibitem{fawzime} G. Belanger, F. Boudjema, F. Donato, R. Godbole and
S. Rosier-Lees, {\bf hep-ph/0002039}.
\bibitem{tevwg} J.F. Gunion et al, {\bf hep-ph/9703330},
in {\em Proccedings of the 1996 DPF/PDB Summer Study of 'New directions 
in high energy physics' Snowmass(1996), Colarado}. 
\bibitem{gunion} J.F. Gunion, {\bf hep-ph/9705282} in  Ref.~\protect\cite{5}.
\bibitem{abdel3}A. Djouadi, {\bf hep-ph 9910449}.
\bibitem{abdel4} See for example, A. Djouadi, W. Kilian, M. Muhlleitner and
P.M. Zerwas, {\bf hep-ph/0001169}.
\bibitem{AD1} A.Datta,M. Guchait and M. Drees, \Journal{\ZPC}{69}{347}{1996}
({\bf hep-ph 9503431}).
\bibitem{ambrasanio} For example, see, S. Ambrosanio,  in 
Ref.~\protect\cite{6P}; H. Baer, P.G. Mercadante, X. Tata and Yili Wang,
{\bf hep-ph 0004001}. 
\bibitem{mura} See for example, A. Gouvea, A. Friedland and H. Murayama,
\Journal{\PRD}{59}{095008}{1999}({\bf hep-ph/9803481}). 
\bibitem{krasnikov} S.I. Bityukov and N.V. Krasnikov, {\bf hep-ph 9907257,
hep-ph 9903519}.
\bibitem{AD} A.Datta, A.K. Datta and M.K. Parida, \Journal{\PLB}{431}{347}{1998}.
\bibitem{AD2} A.Datta,A.K. Datta. M. Drees and D.P. Roy,\Journal{\PRD}{61}{055003}{2000},
({\bf hep-ph 9907444}).
\bibitem{Xrecent}  H. Baer, M. A. Diaz, P. Quintana and X. Tata,
{\bf hep-ph 0002245}.
\bibitem{xer-tri}H. Baer, M. Drees, F. paige, P. Quintana and X. Tata, {\bf
hep-ph/9906233 v2}.
\bibitem{spira_les} M. Spira in \protect\cite{6P}.
\bibitem{lighttopus} H. Baer, M. Drees, R.M. Godbole, J.F.Gunion and X.Tata, \Journal{PRD}{44}{725}{1991}
\bibitem{abdel6}A. Djouadi, J.-L. Kneur and G. Moultaka, {\bf hep-ph 9910269}.
\bibitem{hinchliff}I. Hinchliffe, F.E. Paige, M.D. Shapiro, J. Soderqvist 
and W. Yao, \Journal{\PRD}{55}{5520}{1997}.
\bibitem{Mihoko1} M.M. Nojiri, {\bf hep-ph 9907530}.
\bibitem{Mihoko2} M.M. Nojiri  and Y. Yamada, \Journal{\PRD}{60}{015006}{1999}.
\bibitem{Mihoko3} D. Toya, T. Kobayashi and M.M. Nojiri, {\bf hep-ph 001267}.
\bibitem{Iashvelli} I. Iashvili and A. Kharchilava, 
\Journal{\NPB}{526}{153}{1998}.
\bibitem{XT1}T. Tsukamoto, K. Fujii, H. Murayama, M.Yamaguchi and Y.Okada,
\Journal{\PRD}{51}{3153}{1995}. 
\bibitem{XT2} J.L. Feng, M.E. Peskin, H. Murayama and X. Tata, 
\Journal{\PRD}{52}{1418}{1995}.
\bibitem{XT3} H. Baer, R. Munroe and X. Tata, \Journal{\PRD}{54}{6735}{1996}.
\bibitem{MP} H. Murayama and M.E. Peskin, {\em Ann. Rev. of Nucl. and Part. 
Sci.} {\bf 46}, 533, 1996.
\bibitem{FM} M.M. Nojiri, K.Fujii and T. Tsukamoto, 
\Journal{\PRD}{54}{6756}{1996}. 
\bibitem{martyn} H.-U. Martyn and G.A. Blair, {\bf hep-ph/9910416}.
\bibitem{ghoshme}  D. K. Ghosh, R.M. Godbole and S. Raychaudhury,
{\bf hep-ph 9904233}.
\bibitem{abdel5} S.Y. Choi, A. Djouadi, M. Guchait,J. Kalinowski, H.S. Song and
P.M. Zerwas,  {\bf hep-ph 0002033}.
\bibitem{man_eboli} M. Drees and O.J.P. Eboli, \Journal{EPJ}{10}{337}{1999}.
\bibitem{sabine} A. Bartl, U.Eberl,S. Kraml,W. Majerotto and W. Porod, 
{\bf hep-ph 0002115} and references therein.
\bibitem{dekg} M. Drees, O.J.P. Eboli, R.M. Godbole and S. Kraml,
in \protect\cite{6P}.
\bibitem{feng_finnel} J.L. Feng and D.E. Finnel, 
\Journal{\PRD}{49}{2369}{1994}. 
\bibitem{others}H-C. Cheng, J.L. Feng and N. Polonsky,
\Journal{\PRD}{56}{6875}{1997}; D {\bf 57}, 152 (1998), 
E. Katz, L. Randall and S. Su, \Journal{\NPB}{536}{3}{1999}.
\end {thebibliography}
\end{document}